\documentclass{article}
\usepackage[utf8]{inputenc}
\usepackage{amsmath}
\usepackage{amssymb}
\usepackage{amsthm}
\usepackage{physics}
\usepackage{dsfont}
\usepackage{bm}
\usepackage{appendix}
\usepackage{comment}
\usepackage{tablefootnote}
\usepackage{empheq}
\usepackage{hyperref}

\usepackage{mathrsfs} 

\title{Augmented Lagrangian method for coupled-cluster}
\author{Fabian M. Faulstich\footnotemark[1], Yuehaw Khoo\footnotemark[2], Kangbo Li\footnotemark[3]}

\usepackage[dvipsnames]{xcolor}
\usepackage{graphicx}
\usepackage{caption}
\usepackage{subcaption}

\newtheorem{theorem}{Theorem}

\newtheorem{proposition}[theorem]{Proposition}

\usepackage{simpler-wick}
\newcommand{\vr}{\mathbf{r}}
\newcommand{\vR}{\mathbf{R}}

\usepackage[symbol]{footmisc}

\begin{document}

\maketitle
\begin{abstract}
   We propose to improve the convergence properties of the single-reference coupled cluster (CC) method through an augmented Lagrangian formalism. The conventional CC method changes a linear high-dimensional eigenvalue problem with exponential size into a problem of determining the roots of a nonlinear system of equations that has a manageable size. However,
    current numerical procedures for solving 
    this system of equations to get the lowest eigenvalue suffer from two practical issues: First,
    solving the CC equations may not converge, and second, when converging, they
    may converge to other -- potentially unphysical --  states, which are stationary points of the CC energy expression. We show that both issues can be dealt with when a suitably defined energy is minimized in addition to solving the original CC equations. We further propose an augmented Lagrangian method for coupled cluster (alm-CC) to solve the resulting constrained optimization problem.     
    We numerically investigate the proposed augmented Lagrangian formulation showing that the convergence towards the ground state is significantly more stable and that the optimization procedure is less susceptible to local minima. Furthermore, the computational cost of alm-CC is comparable to the conventional CC method.
\end{abstract}

\footnotetext[1]{Department of Mathematics, Rensselaer Polytechnic Institute, Troy, NY}
\footnotetext[2]{Department of Computer Science, Cornell University, Ithaca, NY}
\footnotetext[3]{Department of Statistics, The University of Chicago, Chicago, IL}
        
\section{Introduction}

Describing electrons in atoms, molecules and solids is a long-lasting problem in quantum many-body theory. With its diverse applications in chemistry and materials science, the field of {\it electronic structure theory} holds vast implications for the mathematical sciences. Integrating state-of-the-art methods from optimization into this field leads to the development of precise and scalable numerical methods, enabling extensive {\it in silico} studies of chemistry for e.g.~sustainable energy, green catalysis, and nanomaterials. The governing equation that describes such a rich variety of applications is a simple-looking eigenvalue problem of the form 
\begin{equation}
\label{eq:eigenvector}
H \ket{w} = E \ket{w},
\end{equation}
where $H\in \mathbb{C}^{2^K\times 2^K}$ is the matrix (called Hamiltonian), $\ket{w}\in\mathbb{C}^{2^K}$ is the eigenfunction (called wavefunction), and $E$ is the eigenvalue (called energy). The number $K$ is related to the number of basis functions used for discretizing the electronic Schrödinger equation. What prohibits us from solving this eigenvalue problem directly using conventional tools from numerical linear algebra is the scaling of the problem. The dimensionality of $H$ grows exponentially with $K$, which is proportional to the number of electrons. This may not be a problem for the systems considered in this article which are four electron systems leading to $2^8\times 2^8$ matrices, however, for more complex systems we may easily obtain matrices of size $2^{80}\times 2^{80}$ which would na\"ivly require about $10^{22}$ GB to merely store the matrix. We refer readers to the Appendix (Section~\ref{section: background Seq}) for a brief introduction to the electronic Schrödinger equation leading to Eq.~\eqref{eq:eigenvector}.

\subsection{Previous works}
Although the eigenvalue problem Eq.~\eqref{eq:eigenvector} seems to be a daunting task, the computational chemistry and physics community have produced several numerical methods that can tackle problems of such sizes efficiently, balancing the quality of the approximation and the computational cost. In this article, we will study one of the most widely used high-accuracy wavefunction methods in computational chemistry: The coupled cluster (CC) method -- also known as the {\it gold standard} of computational chemistry. From the conventional linear algebra perspective, the coupled cluster method is a strikingly different approach. In a nutshell, it seeks a nonlinear parameterization to represent the eigenvector in Eq.~\eqref{eq:eigenvector}, i.e.,
$$
\ket{w}
=
e^{T(t)}\ket{\phi_0}
$$
where $\ket{\phi_0}\in \mathbb{R}^{2^K}$ is a fixed basis vector and $T(t)\in \mathbb{R}^{2^K \times 2^K}$ is the new unknown parametrized using a comparably small number of parameters $t$.
This ansatz yields a system of low-order polynomial equations -- the CC equations -- where the number of equations scales as the number of parameters.
The roots of this system correspond to parameters that characterize the eigenvectors. While widely successful, the inherent nonlinearity of this polynomial system introduces several computational challenges, distinct from those typically encountered in conventional linear eigenvalue problems. 
In particular, difficulties with convergence may arise when there are several plausible roots~\cite{vcivzek1966correlation,piecuch1994solving,mihalka2023exploring,shavitt2009many} or unstable fixed point~\cite{szakacs2008stability,adams1981symmetry}.

Many works focus on accelerating the convergence. For example, different extrapolation techniques have been proposed, such as the direct inversion in the iterative subspace (DIIS) method~\cite{pulay1980convergence} in the CC framework~\cite{scuseria1986accelerating,ettenhuber2015discarding}. To further speed up the iterative procedure, several approaches have been proposed such as following a different iteration pathway~\cite{matthews2015accelerating} or approximating the Jacobian of the nonlinear equation~\cite{kjonstad2020accelerated,yang2020solving}. More recently, there are approaches based on modifying the Hamiltonian with extra potential, to enhance the convergence to the correct solution~\cite{mihalka2023exploring}.

\subsection{Contribution} 

In this article, we propose a drastically different approach to the aforementioned works. While all previous methods try to address the convergence issues of the conventional CC methods, which are based on solving the CC equations, we address the problem using an optimization framework. We propose to minimize an energy term in addition to solving the CC equations. This changes the problem of determining the roots, to a problem of determining the minimizer. In particular, the extra energy term eliminates the need to consider multiple roots in numerically challenging systems. To solve this constrained optimization problem, we propose an augmented Lagrangian formulation (alm-CC). It consists of an outer loop that adjusts the penalty terms coming from the CC equations and an inner loop minimizing the energy simultaneously with a CC equation penalty term. We provide numerical experiments demonstrating the superior convergence performance of alm-CC with a computational cost that is roughly the same as conventional CC methods.\\

The remaining article is organized as follows: In Section~\ref{sec:Theory} we present and motivate the augmented Lagrangian formulation of the CC approach, christened alm-CC. Section~\ref{sec:Background} provides a detailed background of the underlying matrix structures and a road map to an efficient implementation of the alm-CC method. In Section~\ref{sec:Numerics} we provide numerical investigations using a pilot implementation of the proposed method.

\section{Optimizing coupled cluster}
\label{sec:Theory}
In this section, we proposed an optimization framework as an alternative to the CC method. To that end, we first introduce the conventional CC method. 
The starting point is a particular formulation of the eigenvalue problem:
\begin{equation}\label{eq:general-cc-constraint}
    P_{v,w} H \ket{w} = 0,
\end{equation}
where  $P_{v,w} = I - \ket{w}\bra{v}$, and the column vector $\ket{w}$ and row vector $\bra{v}$ fulfills $\braket{v}{w} = 1$ (this choice of normalization prevents a trivial solution $\ket w = 0$). If we solve Eq.~\eqref{eq:general-cc-constraint} with a weak formulation, i.e. by testing this equation with some choices of $\bra{u}$, we obtain the CC equations $\bra{u}P_{v,w} H \ket{w} =0$. Given a solution $\ket{w}, \bra{v}$ to the CC equations, one can obtain the eigenvalue as 
\begin{equation}
E_{CC}(v,w):=\bra{v} H \ket{w}.
\end{equation}
While $\bra{v}$ can be general as long as it satisfies this normalization constraint, as a special case, if we choose $\bra{v} =
\bra{w}/\braket{w}{w}$, then $P_{v,w}$ reduces to an orthogonal projector, $E_{CC}$ reduces to the Rayleigh quotient, and Eq.~\ref{eq:general-cc-constraint} reduces to a standard eigenvalue equation.

However, the solution to Eq.~\eqref{eq:general-cc-constraint} may or may not be a ground state depending on the initial guess. Furthermore, to avoid the curse-of-dimensionality, CC theory employs a certain nonlinear parameterization for $\ket{w},\bra{v}$ and test functions $\bra{u}$, which further gives rise to potentially approximating a root that does not correspond to the energetically lowest state.
To this end, in addition to solving the CC equation, we minimize the CC energy $E_{CC}$ as well in order to regularize the problem. This change of perspective from root finding to minimizing significantly helps with avoiding spurious roots. 

\subsection{alm-CC}

In this section, we propose an augmented Lagrangian framework for the coupled cluster theory. As mentioned previously, we want to solve the following minimization problem:
\begin{equation}
\label{eq:proj-constraint}
    \min_{\substack{\ket{v}, \ket{w} \in \mathbb{C}^{2^K}\\  
    \braket{v}{w}=1}} E_{CC}(v,w),
    \quad\text{s.t. }\ P_{v,w} H \ket{w} = 0.
\end{equation}
The constraints imply that $\ket{w}$ is an eigenvector of $H$ with $E_{CC}(v,w)$ being its eigenvalue. The solution to Eq.~\eqref{eq:proj-constraint} is the lowest eigenvalue. However, one does not readily get a continuous optimization problem. The issue is that the constraints in Eq.~\eqref{eq:proj-constraint} lead to a measure zero feasible set for $\ket{w}$ that is disconnected, which is the set of eigenvectors with a suitable normalization. To make the optimization continuous, one can in principle relax the hard constraints by adding a soft penalty term. However, a penalty method can suffer from a bad condition number; we thus pursue an augmented Lagrangian method. 

The first step to derive the augmented Lagrangian is to rewrite Eq.~\eqref{eq:proj-constraint} in an equivalent min-max formulation:
\begin{equation}
\label{eq:minmax}
\min_{\substack{\ket{v}, \ket{w} \in \mathbb{C}^{2^K} \\ 
    \braket{v}{w}=1}} 
\max_{\ket{u} \in \mathbb{C}^{2^K}}  \bra{v} H \ket{w} + \bra{u} H \ket{w} - \braket{u}{w}\bra{v} H \ket{w}.
\end{equation}
Then one can alternate between the minimization and the maximization. In this min-max formulation, the hard constraints are relaxed, through a Lagrange multiplier $\ket{u}$, and now it is possible to optimize $\ket{w}$ continuously. 

To address the exponentially large vectors $\ket w, \ket v$, and $\ket u$ in Eq.~\eqref{eq:minmax}, we parameterize $\ket w, \ket v,\ket u$ as $\ket {w(t)}, \ket{v(t)},\ket{u(t)}$, respectively, where $t,\;\lambda \in\mathcal{V}$ with  $\mathcal{V}$ being a low-dimensional vector space (vide infra). This yields the parameterized min-max problem \begin{equation}
\label{eq:minmax in t lambda}
\min_{t\in \mathcal{V}} \max_{\lambda \in \mathcal{V}} \bra{v(t)} H \ket{w(t)} + \bra{u(\lambda)} H \ket{w(t)} - \braket{u(\lambda)}{w(t)}\bra{v(\lambda)} H \ket{w(t)}.
\end{equation}
While the decision variable $t$ is small, the loss function can still be expensive to evaluate, as it requires taking an inner product of exponentially sized vectors. Following CC theory, we use an exponential ansatz for the state vectors~\cite{hubbard1957description,hugenholtz1957perturbation}:
\begin{align}
\label{eq:cc-param}
\ket{w(t)} = e^{T(t)}\ket{\phi_0}, \quad
\bra{v(t)} = \bra{\phi_0} e^{-T(t)}, \quad
\bra{u(\lambda)} = \bra{\phi_0} \Lambda(\lambda) e^{-T(t)},
\end{align}
with $\braket{\phi_0}{\phi_0} = 1$. In Eq.~\eqref{eq:cc-param}, $\Lambda(\lambda)$, $T(t) \in \mathbb{C}^{2^K \times 2^K}$ are cluster matrices that we will introduce in detail in Section~\ref{sec:Background}, along with the efficient evaluation of the Hamiltonian matrix elements. 
This parametrization also naturally enforces $\braket{v(t)}{w(t)}=1$, and  $\braket{u(\lambda)}{w(t)} = 0$ (the latter equality simplifies the expressions in Eq.~\eqref{eq:minmax in t lambda}). Substituting Eq.~\eqref{eq:cc-param} 
into Eq.~\eqref{eq:minmax in t lambda} yields
\begin{equation}
\label{eq:minmax_t_lambda cc}
\min_{t \in \mathcal{V}} \max_{\lambda\in \mathcal{V}} 
\bra{ \phi_0} (I+\Lambda(\lambda)) e^{-T(t)} H e^{T(t)} \ket{\phi_0}.
\end{equation}
Although it is not obvious that the loss in Eq.~\eqref{eq:minmax_t_lambda cc} can be evaluated efficiently, we will show in Section~\ref{sec:Background} how this can be done using the algebra of the exponential ansatz developed in CC theory. 

A remaining problem is that the parametrization of $\Lambda(\lambda)$ in Eq.~\eqref{eq:cc-param} is linear w.r.t.~$\lambda$, hence, the maximization over $\lambda$ in Eq.~\eqref{eq:minmax_t_lambda cc} is unbounded. This may create difficulties if we want to alternate between minimizing and maximizing in  Eq.~\eqref{eq:minmax_t_lambda cc}. We therefore introduce a proximal point term~\cite{nocedal2006quasi}, penalizing the deviation from the previous estimate of $\lambda$ denoted by $\tilde \lambda$.
This yields
\begin{equation}
\begin{aligned}
\min_{t\in \mathcal{V}} \max_{\lambda\in \mathcal{V}}
\bra{\phi_0} (I + \Lambda(\lambda)) e^{-T(t)}H e^{T(t)} \ket{\phi_0 }
-\frac{1}{ 2\alpha} |\lambda - \tilde \lambda |^2.
\end{aligned}
\end{equation}
Maximizing w.r.t.~$\lambda$ then yields
\begin{equation}
\lambda
=
\tilde\lambda + \alpha g(t), \quad \text{where }
g_\mu(t)
=
\bra{ \phi_\mu} e^{-T(t)}H e^{T(t)} \ket{ \phi_0}.
\end{equation}
Then we find the augmented Lagrangian
\begin{equation}
\begin{aligned}
\mathcal{L}_\alpha(t, \tilde \lambda)
&=
\max_{\lambda}~
\bra{\phi_0} (I + \Lambda(\lambda)) e^{-T(t)}H e^{T(t)} \ket{\phi_0}
-\frac{1}{ 2\alpha} |\lambda - \tilde \lambda |^2\\
&=
\bra{ \phi_0} (I +\Lambda(\tilde \lambda)) e^{-T(t)}H e^{T(t)} \ket{\phi_0 }
+ \alpha |g(t)|^2
-\frac{\alpha}{ 2} | g(t) |^2\\
&=
\bra{ \phi_0} (I +\Lambda(\tilde \lambda)) e^{-T(t)}H e^{T(t)} \ket{\phi_0}
+\frac{\alpha}{ 2} | g(t) |^2,
\end{aligned}
\end{equation}
which can be rewritten as 
\begin{equation}
\begin{aligned}
\mathcal{L}_{\alpha} (t, \lambda)
&= 
E(t) 
+ \sum_{\mu} \lambda_{\mu} g_{\mu}(t) 
+ \frac{\alpha}{2} \sum_{\mu} \norm{g_{\mu}(t)}^2.
\end{aligned}
\end{equation}
The gradient of the augmented Lagrangian then takes the form
\begin{equation}
\frac{\delta \mathcal{L}_{\alpha} (t, \lambda)}{\delta t_{\nu}} 
= 
\frac{\delta E}{ \delta t_{\mu}}
+ \sum_{\mu} \lambda_{\mu} \frac{\delta g_{\mu}}{\delta t_{\nu}} 
+ \alpha \sum_{\mu} \frac{\delta g_{\mu}}{\delta t_{\nu}} g_{\mu},
\end{equation}
where the individual derivatives are 
\begin{align}
\frac{\delta E}{\delta t_{\mu}}  = 
\bra{\phi_0} H e^{T(t)} X_{\mu} \ket{\phi_0},\quad
\frac{\delta g_{\mu}}{\delta t_{\nu}}  =  
\bra{ \phi_{\mu}} e^{-T(t)} [H, X_{\nu}] e^{T(t)} \ket{\phi_0},\quad
X_{\mu} = \frac{T(\delta t)}{\delta t_{\mu}}.
\label{eq:derivativeT}
\end{align}
The evaluation of the gradient is asymptotically at most as expensive as the objective function itself and the additional $X_{\mu}$ does not add to the complexity. 
The remaining question is the efficient evaluation of the
matrix elements of the Hamiltonian $\bra{\phi_0} H e^{T(t)}\ket{\phi_0}$ and
$\bra{\phi_0} \Lambda(\lambda) e^{-T(t)} H e^{T(t)}\ket{\phi_0}$, which we discuss in the following section.

\section{The cluster matrices}
\label{sec:Background}

To understand the mathematical structure of the CC ansatz, it is necessary to
introduce the fermionic creation and annihilation matrices, which the CC ansatz
uses to satisfy the anti-symmetric constraints arising from the spin-statistics
theorem. There are other considerations that makes the CC ansatz effective in
the sense that it compactly captures the space of low energy states. 
However, the rationale behind those design choices are not necessary for
explaining what the CC ansatz is. Instead,
in this section, we aim to present a concise mathematical exploration of these matrix structures, delving into the CC method and providing a road map to an efficient implementation of the alm-CC method. For further details, we refer the reader to~\cite{faulstich2024recent,helgaker2013molecular,shavitt2009many}.

\subsection{Underlying matrix structures}
\label{sec:SecondQuant}

We begin this section by introducing the Galekin space used for the discretization of the continuous electronic Schrödinger equation, see $\mathscr{H}$ in Eq.~\eqref{eq:ElecSE}.
To that end, let $\mathcal{B}$ be an $L^2$-orthonormal basis of a finite-dimensional Hilbert space $h\subset H^1(X)$, where $X=\mathbb{R}^3 \times \{ \pm 1/2 \}$ and $|\mathcal{B}|=K\gg N$ with $N$ being the number of electrons in the system.
As fermions, electrons must adhere to the Pauli exclusion principle, which is reflected in the anti-symmetry of the eigenfunctions~\cite{yserentant2010regularity}. 
We therefore employ an anti-symmetric product ansatz: We refer to the $M$-folded exterior power of $h$ as the $M$-particle Hilbert space equipped with the natural orthonormal basis, denoted $\mathfrak{B}^{(M)}$, coming from $M$-folded wedge-products of elements of $\mathcal{B}$.
Next, we define the Galerkin space 
\begin{equation}
\label{eq:FockSpace}
\mathcal{F} = \bigoplus_{M=0}^{K} \mathcal{H}^{(M)},
\end{equation}
which is used to discretize the electronic structure Hamiltonian in Eq.~\eqref{eq:ElecSE}.
For brevity, we will write the basis elements in $\mathcal{F}$ as
\begin{equation}
\ket{s_1,...,s_{K}} = \frac{1}{\sqrt{M!}}\chi_1^{s_1} \wedge \chi_2^{s_2} \wedge ... \wedge \chi_{K}^{s_{K}},
\end{equation}
where $s_I\in \{0,1\}$ for all $I=1,...,K$, $M = \sum_{i=1}^{K} s_i$ and $\{ \chi_I\}$ is the basis of $h$.
With respect to the natural extension of the $L^2$-inner product onto $\mathcal{F}$, the basis elements in $\mathcal{F}$ are orthonormal.
A general element in $\mathcal{F}$, is then given as 
\begin{equation}
\ket{w}
=
\sum_{s_1,...,s_{K} \in \{ 0, 1\}}
w(s_1,...,s_{K}) \ket{s_1,...,s_{K}},
\end{equation}
where $w(s_1,...,s_{K}) \in \mathbb{C}$.

We now proceed to define the fermionic creation and annihilation matrices used in our analysis. 
These matrices describe the process of adding/removing a particle to/from a quantum state and are denoted by ``$a^\dag$'' and ``$a$'', respectively.
These matrices eliminate the anti-symmetry and normalization constraints.
Let $P\in[\![K]\!]$, we define
\begin{equation}
\begin{aligned}
a_P^\dag: \mathcal{F} \to \mathcal{F} ~;~ 
\ket{s_1,...,s_{K}}
&\mapsto (-1)^{\sigma(p)} (1-s_P) \ket{s_1,...s_{P-1}, 1-s_P, s_{P+1},...,s_{K}}\\
a_P: \mathcal{F} \to \mathcal{F} ~;~\ket{s_1,...,s_{K}}
&\mapsto (-1)^{\sigma(p)} s_P \ket{s_1,...s_{P-1}, 1-s_P, s_{P+1},...,s_{K}}
\end{aligned}
\end{equation}
where $\sigma(p) = \sum_{Q=1}^{P-1} s_Q$. 
Mathematically, the algebra of these matrices follows the anti-commutation relations (CAR) characteristic of fermions, i.e.,
\begin{equation}
[a_P, a_Q]_+ = [a_P^\dag, a_Q^\dag]_+ = 0 \quad \text{and} \quad  [a_P^\dagger, a_Q]_+ = \delta_{p,q}\quad \forall ~1\leq P,Q\leq K,
\end{equation}
capturing the essence of the Pauli exclusion principle. They are key to constructing the Galerkin space representation of $\mathscr{H}$, allowing us to express the many-body Hamiltonian and cluster matrices in a form that is amenable to computational analysis.
Note that by construction $\mathcal{F} \simeq \mathbb{C}^{2^{K}}$, hence, the basis elements of $\mathcal{F}$ can be expressed as
\begin{equation}
\ket{s_1,...,s_{K}}
=
{ 1 - s_1 \choose s_1 } \otimes ... \otimes { 1 - s_{K} \choose s_{K} } .
\end{equation}
Then, the fermionic creation and annihilation operators are sparse matrices of the form
\begin{equation}
\begin{aligned}
a_P^\dagger
=
\underbrace{\sigma_z \otimes ... \otimes \sigma_z}_{P-1~{\rm times}} 
\otimes\;
a^\dagger
\otimes 
\underbrace{I \otimes ... \otimes I}_{K-P-1~{\rm times}}
\quad 
{\rm and}
\quad
a_P
=
\underbrace{\sigma_z \otimes ... \otimes \sigma_z}_{P-1~{\rm times}} 
\otimes\;
a
\otimes 
\underbrace{I \otimes ... \otimes I}_{K-P-1~{\rm times}}
\end{aligned}
\end{equation}
where 
\begin{equation}
I = 
\begin{pmatrix}
1 & 0 \\
0 & 1
\end{pmatrix},\quad
\sigma_z = 
\begin{pmatrix}
1 & 0 \\
0 & -1
\end{pmatrix},\quad
a = 
\begin{pmatrix}
0 & 1 \\
0 & 0
\end{pmatrix}.
\end{equation}
Note that this implies directly $(a_P^\dagger)^2 = (a_P)^2 = 0$ for all $1\leq P \leq K$.

In this formulation, the matrix describing an interacting electronic system in a potential generated by clamped nuclei, i.e., the electronic structure Hamiltonian, takes the form
\begin{equation}
\label{eq:ElecHam}
H
=
\sum_{P,Q} h_{P,Q} a_P^\dag a_Q
+
\frac{1}{4}
\sum_{P,Q,R,S} v_{P,Q,R,S} a_P^\dag a_Q^\dag a_S a_R,
\end{equation}
where $h\in\mathbb{C}^{K \times K}$ and $v \in \mathbb{C}^{K \times K \times K \times K}$ are system dependent~\cite{helgaker2013molecular}.

\subsection{The coupled cluster ansatz}
\label{sec:CCansatz}

The efficiency of the CC theory is achieved by parametrizing the wavefunction in 
$\mathcal{H}^{(N)}$ as a rapidly converging series. The starting point of the ansatz is a 
\textit{reference vector}
\begin{equation}
\ket{\phi_0} = a^\dagger_N....a^\dagger_1 \ket{0,...,0} \in \mathcal{H}^{N},
\end{equation}
where the first $N$ entries are set to one, and the remaining entries are zero. Coupled cluster theory is built on the exponential ansatz, i.e.,
\begin{equation}
\label{eq:ExpAnsatz}
\ket{w}
=
\exp(T) \ket{\phi_0},
\end{equation}
where $T\in \mathbb{R}^{ 2^{K} \times 2^{K}}$ is the new unknown.
To determine the parameters of the ansatz, i.e., the matrix $T$, one substitutes the ansatz in 
Eq.~\eqref{eq:ExpAnsatz} into Eq.~\eqref{eq:eigenvector} which yields
\begin{equation}
\label{eq:CC}
H \exp(T) \ket{\phi_0}
= E \exp(T) \ket{\phi_0}
\quad  \Leftrightarrow  \quad 
\left\lbrace
\begin{aligned}
\bra{\phi_0} \exp(-T) H \exp(T) \ket{\phi_0}
&= E,\\
P(\ket{\phi_0}) \exp(-T) H \exp(T) \ket{\phi_0}
&= 0,  \\
\end{aligned}
\right.
\end{equation}
where we introduced the convention that $\bra{w} = \ket{w}^*$ is the adjoint. 
The second set of equations 
\begin{equation}
\label{eq:CCEqs}
P(\ket{\phi_0}) \exp(-T) H \exp(T) \ket{\phi_0}
= 0,
\end{equation}
is a system of nonlinear equations -- known as the CC equations -- determining $T$. 
We shall now elaborate on the matrix space over which we seek to find $T$, which we call the {\it space of cluster matrices}. 
We begin by introducing the excitation matrices: Let $A_1,...,A_k \in v_{\rm virt}$, and $I_1,...,I_k \in v_{\rm occ}$, where
$v_{\rm occ} = [\![ N ]\!]$ and $v_{\rm virt} =[\![ K ]\!] \setminus [\![ N ]\!]$.
We define the excitation matrices 
as
\begin{equation}
X_{A_1,...,A_k \choose I_1,...,I_k}
:=
a_{A_k}^\dagger ... a_{A_1}^\dagger a_{I_k} ... a_{I_1},
\end{equation}
and the set of all excitation matrices is then given by
\begin{equation}
\mathfrak{E}(\mathcal{H}^N)
:=
\left\lbrace 
X_\mu ~\Big|~ \mu = {A_1,...,A_k \choose I_1,...,I_k},\, A_j \in  v_{\rm virt}, \, I_j \in v_{\rm occ},\, k\leq N
\right\rbrace.
\end{equation}
The excitation matrices then span the space of cluster matrices  
\begin{equation}
\label{eq:expansion}
\mathfrak{b}
=
{\rm span}(\mathfrak{E}(\mathcal{H}^N))
=
\left\lbrace
T~\Bigg|~
T=\sum_\mu t_\mu X_\mu
\right\rbrace
.
\end{equation} 
In the CC method, we seek to find the unknown $T$ within $\mathfrak{b}$ or a suitable subspace. 
Central to the numerical feasibility of the exponential parametrization are two theorems: 
The first is that the excitation matrices commute, which dramatically simplifies the algebra, especially when taking the gradient.
\begin{proposition}
\label{th:Commutator}
Let $X_\mu, X_\nu \in \mathfrak{E}(\mathcal{H}^N)$. Then
$
[X_\mu, X_\nu] = 0
$.
\end{proposition}
We provide the proof in the appendix, which follows similar ideas as outlined in~\cite{faulstich2024coupled}. 
The second is that the excitation matrices are nilpotent to order two.
\begin{proposition}
\label{th:nilpotency}
Let $X_\mu \in \mathfrak{E}(\mathcal{H}^N)$. Then $X_\mu^2 = 0$.
\end{proposition}
Note that the matrices in $\mathfrak{b}$, are uniquely defined via the coefficient vector ${\bf t} = (t_\mu)$ which is called the cluster amplitude vector. 
We note that the nilpotency from the excitation matrices carries on to the cluster matrices. 
To see this, we may expand $T^{N+1}(t)$ which yields
\begin{equation}
T^{N+1}(t)
=
\sum_{
\substack{
k_{1}+k_{2}+\cdots +k_{m}=N+1\\ 
k_{1},k_{2},\cdots ,k_{m}\geq 0}}
{N+1 \choose k_{1},k_{2},\ldots ,k_{m}}\prod _{j=1}^{m}(t_{\mu_j} X_{\mu_j})^{k_j}.
\end{equation}
Since $|v_{\rm occ}| = N$, there exists at least one $i\in v_{\rm occ}$ that appears at least twice in the $\prod _{j=1}^{m} X_{\mu_j}^{k_j}$. 
However, since $a_i^2 = 0 $ this yields that $T^{N+1}=0$.  
Hence, the exponential series $\exp(T)$ for any element $T \in \mathfrak{b}$ terminates after at most $N$ terms.
With these results in place, it is possible to show that the CC ansatz is well-defined~\cite{faulstich2024coupled}, i.e., considering any function $\ket{w} \in \mathcal{H}^{(N)}$ fulfilling $\braket{w}{\phi_0} = 1$ there exists a unique cluster amplitude vector such that 
\begin{equation}
\ket{w}
=
\exp(T(t)) \ket{\phi_0}.
\end{equation}
In practice, the expansion of the cluster matrices is commonly truncated to the single and double excitations, i.e., in the expansion in Eq.~\eqref{eq:expansion} we only include of multi indices of the form
\begin{equation}
    \mu = \begin{cases}
        { A \choose I} & \text{singles, i.e., } |\mu| =1,\\
        { {A_1, A_2} \choose {I_1, I_2}} & \text{doubles, i.e., } |\mu| =2.
    \end{cases} 
\end{equation}
Similar to the cluster matrices, we moreover define the adjoint cluster matrices, as 
\begin{equation}
\Lambda(\lambda)
=
\sum_\mu \lambda_\mu X_\mu^\dagger
\end{equation}
where $\lambda = (\lambda_\mu)$ is the {\it lambda amplitude vector}.

\section{Numerics}
\label{sec:Numerics}

In this section, we compare the conventional coupled cluster singles and doubles (CCSD) method -- which is optimized with the traditional (quasi) Newton method -- with the here proposed augmented Lagrangian formulation. Our discussion begins by comparing the two methods for scenarios where CCSD is well-known to provide reliable results. We show that the alm-CCSD method reproduces the CCSD results. We then investigate the performance of alm-CCSD for systems where the conventional CCSD method can become problematic to converge.

\subsection{Agreement with quasi-Newton method}
We begin the numerical exposition of the proposed method by investigating the agreement of the results in scenarios where CCSD is known to work well. To that end, we compare CCSD with alm-CCSD for the helium atom, dissociation of the hydrogen molecule, as well as dissociation of H$_4$ in D$_{2{\rm h}}$ and D$_{\infty {\rm h}}$ configuration.

\begin{figure}[h!]
    \centering
    \includegraphics[width = 0.5\textwidth]{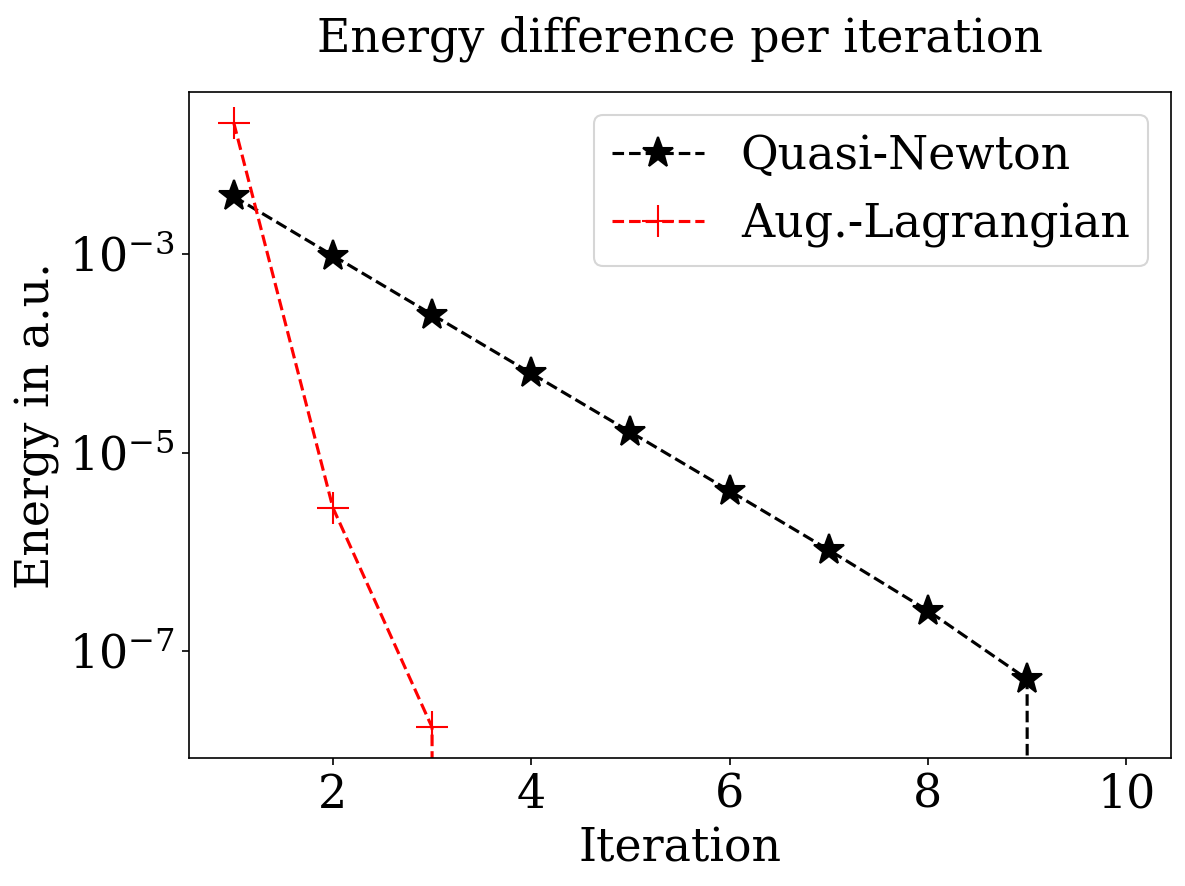}
    \caption{Comparison of the energy error from quasi-Newton CCSD and the outer iterations of alm-CCSD.}
    \label{fig:HeliuOuterConv}
\end{figure}

For the helium atom, we observe that alm-CCSD converges just as fast as the conventional CCSD method. We report the outer iterations of alm-CCSD and the total number of iterations for the quasi-Newton optimization in Fig.~\ref{fig:HeliuOuterConv}. 

Taking the number of inner iterations for the alm optimization into account, we find that alm-CCSD requires just as many total iterations as CCSD. 
The number of inner iterations per outer iteration for the alm-CCSD method are listed in Tab.~\ref{tab:HeInnerIts} 

\begin{table}[h!]
    \centering
            \begin{tabular}{r|cccc}
                 Outer iteration    & 1 & 2 & 3 &4  \\
                 \hline
                 Number of inner iterations &  3 & 4 & 1 & 0
            \end{tabular}
    \caption{Number of inner iterations of alm-CCSD.}
    \label{tab:HeInnerIts}
\end{table}

\begin{figure*}[h]
    \centering
    \begin{subfigure}[t]{0.31\textwidth}
        \centering
        \includegraphics[width = .5\textwidth]{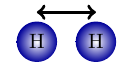}
        \caption{H$_2$}
    \end{subfigure}%
    \begin{subfigure}[t]{0.31\textwidth}
        \centering
        \includegraphics[width = \textwidth]{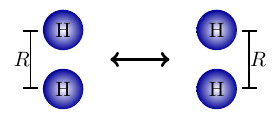}
        \caption{H$_4$ in D$_{2 {\rm h}}$ configuration}
    \end{subfigure}
    \hfill
    \begin{subfigure}[t]{0.31\textwidth}
        \centering
        \includegraphics[width = \textwidth]{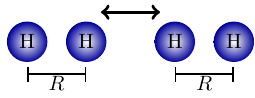}
        \caption{H$_2$ in D$_{\infty {\rm h}}$ configuration}
    \end{subfigure}
        \caption{\label{fig:sys} Schematic depiction of the different dissociation processes. The parameter $R = 1.4$ a.u. which is close to the equilibrium of H$_2$.}
\end{figure*}

Next, we will compare the energies obtained with alm-CCSD to the energies obtained with conventionally optimized CCSD for dissociation processes where the latter is known to yield the correct solution. We compute dissociation curves for H$_2$, and H$_4$ in D$_{2{\rm h}}$ and D$_{\infty {\rm h}}$ configuration, see Fig.~\ref{fig:sys}. 

\begin{figure*}[h]
    \centering
    \begin{subfigure}[t]{0.33\textwidth}
        \centering
        \includegraphics[width = \textwidth]{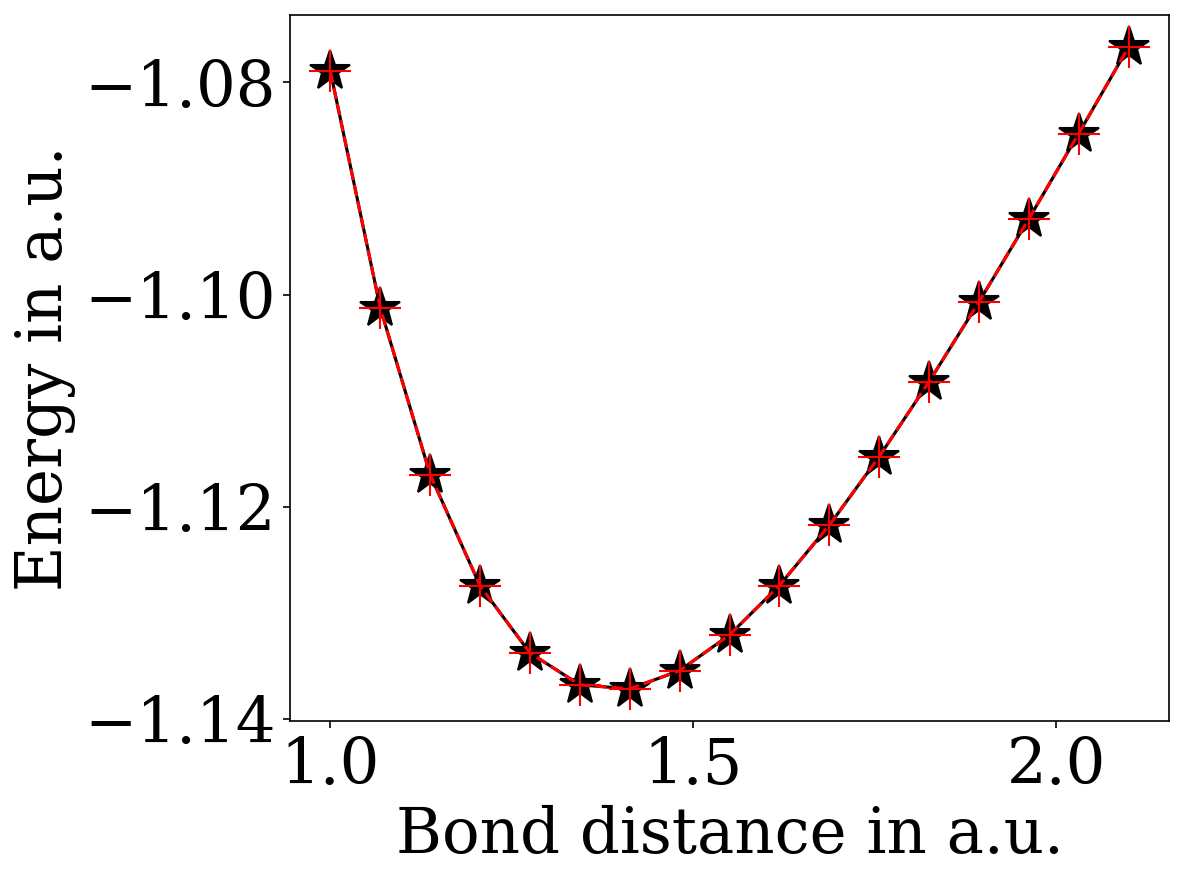}
        \caption{}
    \end{subfigure}%
    \begin{subfigure}[t]{0.33\textwidth}
        \centering
        \includegraphics[width = \textwidth]{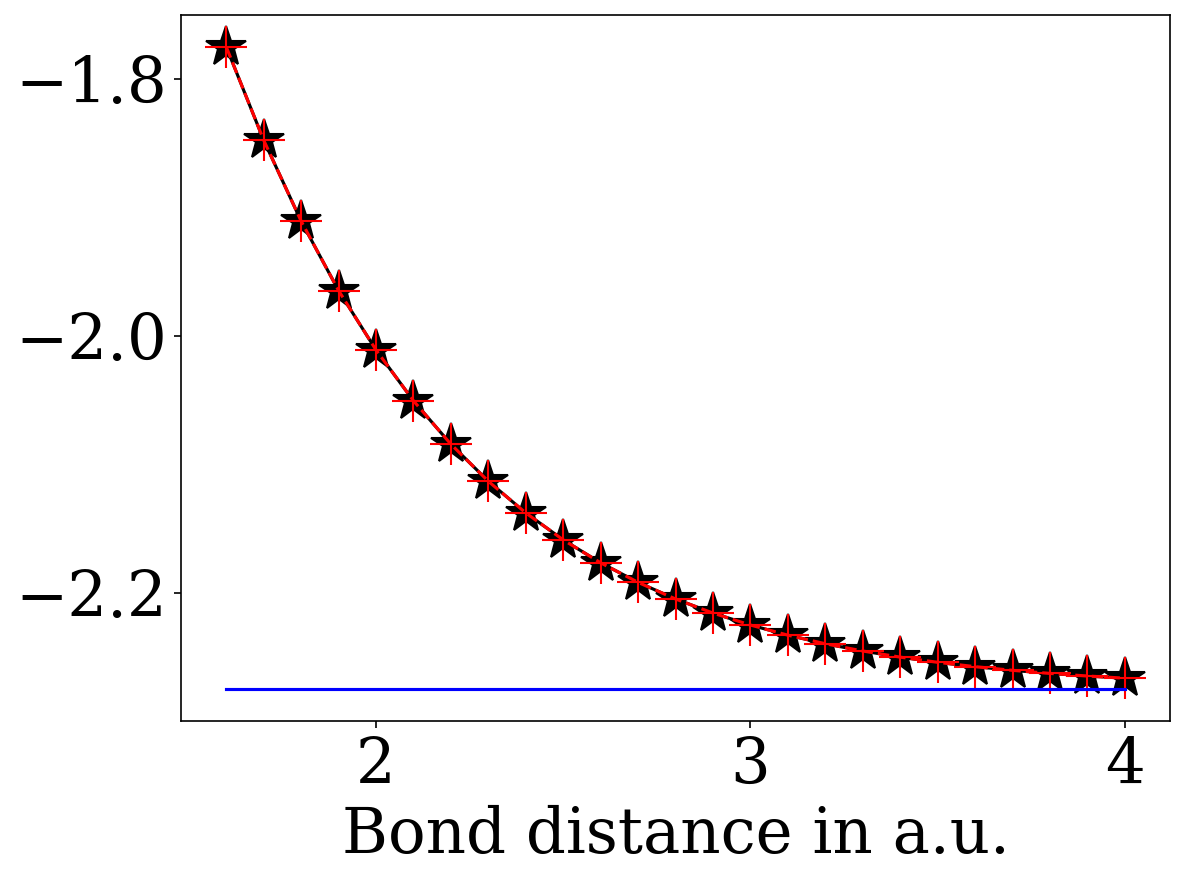}
        \caption{}
    \end{subfigure}
    \begin{subfigure}[t]{0.33\textwidth}
        \centering
        \includegraphics[width = \textwidth]{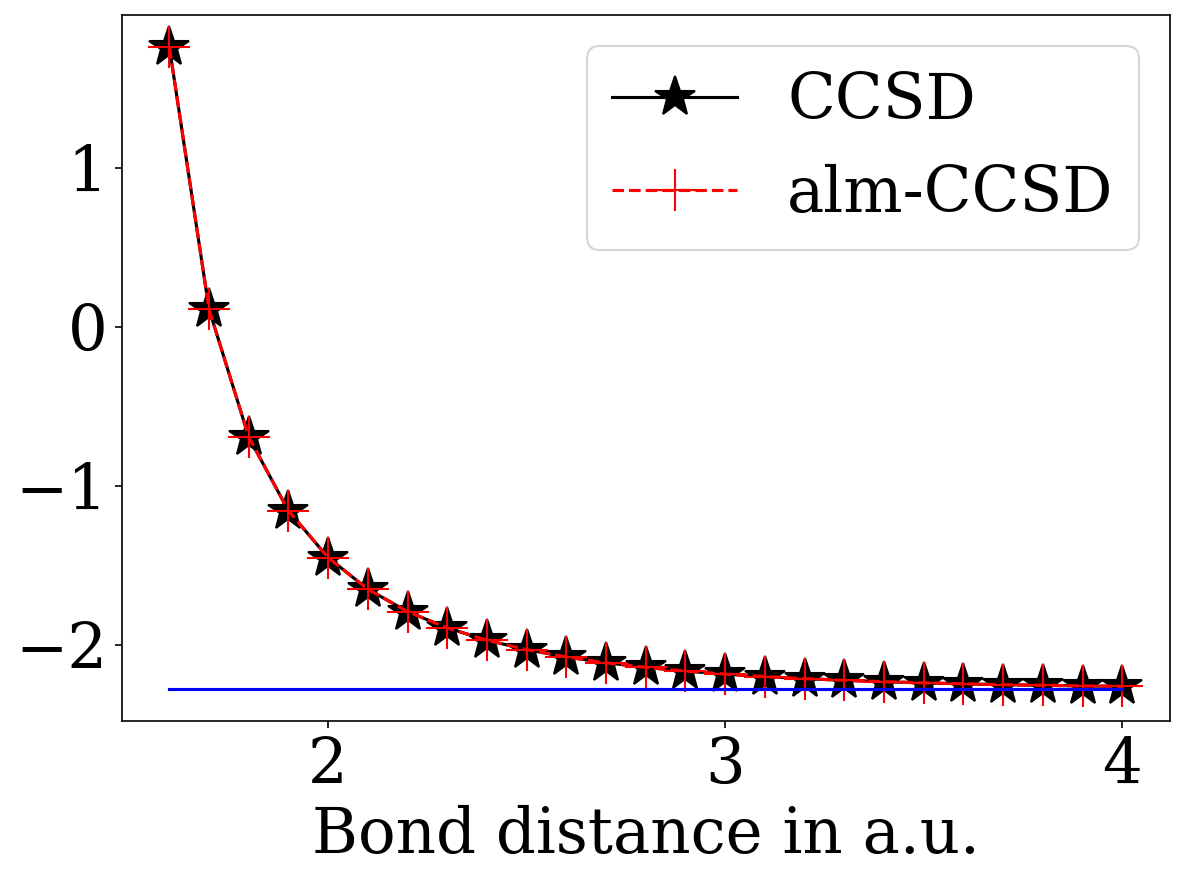}
        \caption{}
    \end{subfigure}
    \caption{\label{PECconparison}Comparison of energies obtained with alm-ccsd and conventional quasi-newton CCSD. In panels (b) and (c) we included the dissociation energy of the respective H$_4$ molecules as a solid blue line, showing the size consistency of the method.}
\end{figure*}

\noindent
For all considered dissociations, we find that alm-CCSD recovers the potential energy curve perfectly, see Fig.~\ref{PECconparison}.

\subsection{Global convergence behavior}
\label{sec:GlobalConvergence}

\begin{figure}[h!]
    \centering
    \includegraphics{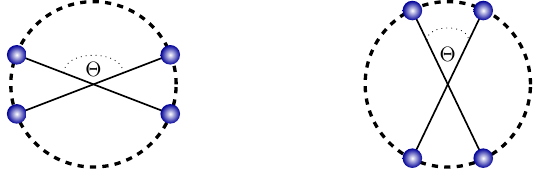}
    \caption{Depiction of the H$_4$ model undergoing a symmetric disturbance on a circle modeled by the angle  $\Theta$.}
    \label{fig:H4_circ_mod}
\end{figure}

To exemplify the advantage of an augmented Lagrangian-based optimization, we now consider a variant of the H$_4$ model consisting of four hydrogen atoms symmetrically distributed on a circle of radius 
$R = 1.738$~\AA~\cite{van2000benchmark}, and follow the system's geometry configuration outlined in~\cite{bulik2015can}, see Fig.~\ref{fig:H4_circ_mod}.

This model system provides rich possibilities for the comparison of the augmented Lagrangian method with the conventional (quasi) Newton-based optimization. 

\subsubsection{Multiple roots}
In this section, we investigate the convergence to an unphysical or unfavorable root -- as a system of polynomial equations, the CC equations have numerous roots, some of which are physical, others are not~\cite{faulstich2023algebraic}.
Depending on the initialization, different roots may be found using a (quasi) Newton approach. 
Stably converging to the energetically lowest root is consequently of great interest to the community. 

\begin{figure}[ht!]
    \centering
    \includegraphics[width = 0.5 \textwidth]{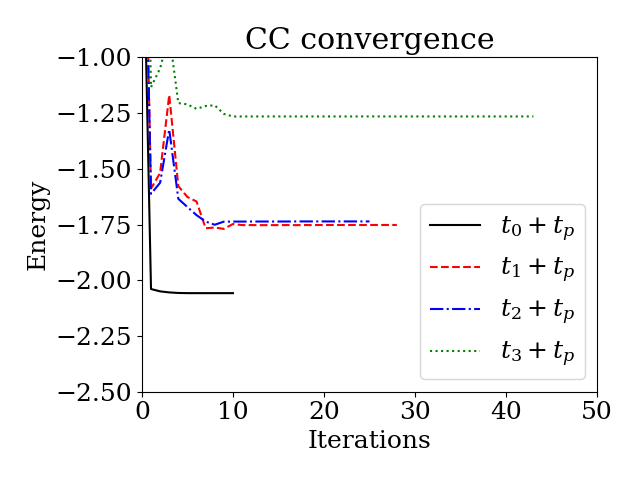}
    \caption{Energy progress from quasi Newton using a random initialization 
    ${\bf t}_i + {\bf t}_p$ with $\Vert {\bf t}_p \Vert/\Vert {\bf t}_i \Vert = 0.1$ 
    for $i = 0, \ldots, 3$.
    }
    \label{fig:CCSD_energy_iterations}
\end{figure}

Using the H$_4$ model with $\Theta = 45^\circ$ (which corresponds to two well-separated hydrogen molecules), we can generate four distinct solutions towards which the (quasi) Newton method stability converges, depending on their initialization.
These four distinct roots ${\bf t}_0$, ${\bf t}_1$, ${\bf t}_2$,  ${\bf t}_3$ yield the respective energies
$$
E_0 = -2.05783;\qquad
E_1 = -1.75205;\qquad
E_2 = -1.73608;\qquad
E_3 = -1.26518.
$$
Clearly, $E_0$ is the energetically lowest one and the closest to the energy obtained by diagonalizing the Hamiltonian, i.e., $E_* = -2.05783$. When initializing the (quasi) Newton method with ${\bf t}_0$, ${\bf t}_1$, ${\bf t}_2$, ${\bf t}_3$ or variants of these amplitude vectors
randomly perturbed by ${\bf t}_p$, it rapidly converges to the respective root, see Fig.~\ref{fig:CCSD_energy_iterations}.

\begin{figure}[ht!]
    \centering
    \includegraphics[width = 0.6 \textwidth]{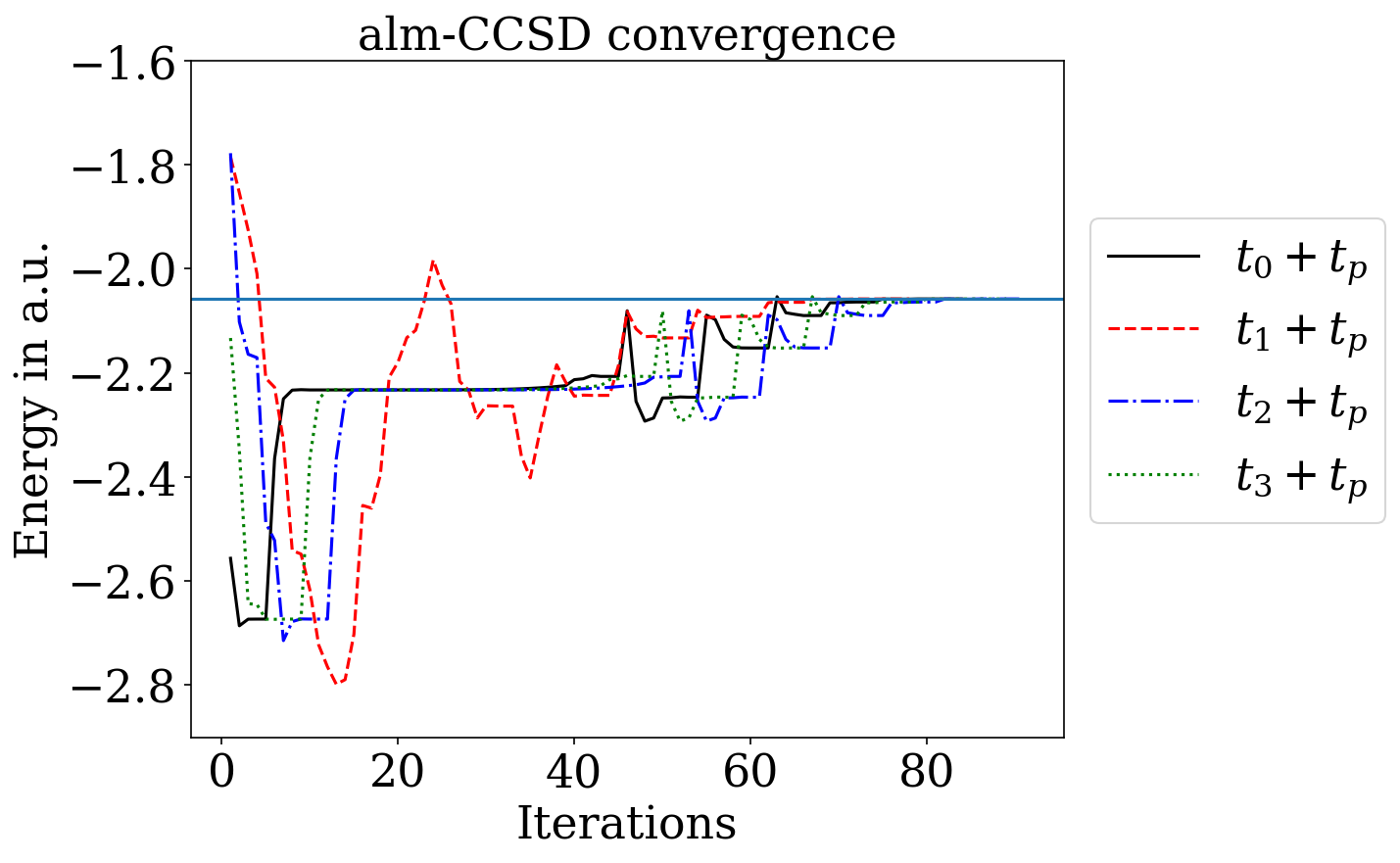}
    \caption{Energy progress from the augmented Lagrangian formulation starting from ${\bf t}_i + {\bf t}_p$ with $\Vert {\bf t}_p \Vert/\Vert {\bf t}_i \Vert = 0.1$. The solid blue line shows that targeted energy $E_0$.}
    \label{tab:aug_CCSD_energy_iterations}
\end{figure}

On the contrary, the augmented Lagrangian formulation stability converges to the ground state ${\bf t}_0$, as shown in Fig.~\ref{tab:aug_CCSD_energy_iterations}. We emphasize that even in the limit $\Vert {\bf t}_p \Vert \to 0$, the augmented Lagrangian formulation yields convergence to ${\bf t}_0$. Although the number of iterations has increased over the conventional (quasi) Newton optimization, this can be attributed to ${\bf t}_0$ being further away than ${\bf t}_i$ from ${\bf t}_i + {\bf t}_p$. Note also, that the convergence behavior of alm-CC can be improved when the method's hyperparameters are optimized. However, as the primary objective of this paper is to present a conceptual improvement over the conventional quasi-newton-based optimization procedure in CC theory through the implementation of an augmented Lagrangian formulation, we defer the optimization of these hyperparameters, along with more comprehensive benchmarking, to future research works.

\subsubsection{Enhanced convergence}
\label{Sec:EnsuringConvergence}

In scenarios where the optimization landscape is more complicated, Newton's method may fail to converge altogether. In this case, multiple parameter adjustments and chemically motivated numerical ``tricks'' can be applied to eventually obtain convergence, yet, the out-of-the-box use of the (quasi) Newton method to converge the CC roots is not possible for such systems. 
In the case of the H$_4$ model, we can observe this at $\Theta = 90^\circ$. Here we can obtain a solution denoted by ${\bf t}_*$ using PySCF with an initial guess ${\bf t}_{\rm init} = 0$. We then add a random perturbation ${\bf t}_p$ to this solution and use it as a different initialization ${\bf t}_{\rm init} = {\bf t}_* + {\bf t}_p$ for the (quasi) Newton method. 
Scaling the size of ${\bf t}_p$, i.e., $\Vert {\bf t}_p \Vert$, allows us to (approximately) investigate the basin of attraction. 
Taking into consideration that the different regions of attraction for Newton's method can yield fractal structures, we expect that the region for which the (quasi) Newton method converges to ${\bf t}_*$ will not be circular. 
However, the performed investigation yields a ballpark for the basin of convergence, since it provides the radius $r_{\rm max}$ of the largest ball in which the (quasi) Newton method converges to the solution ${\bf t}_*$ in $99.99\%$ of the performed test cases. 

\begin{figure}[h!]
     \begin{subfigure}[t]{0.495\textwidth}
    \includegraphics[width = \textwidth]{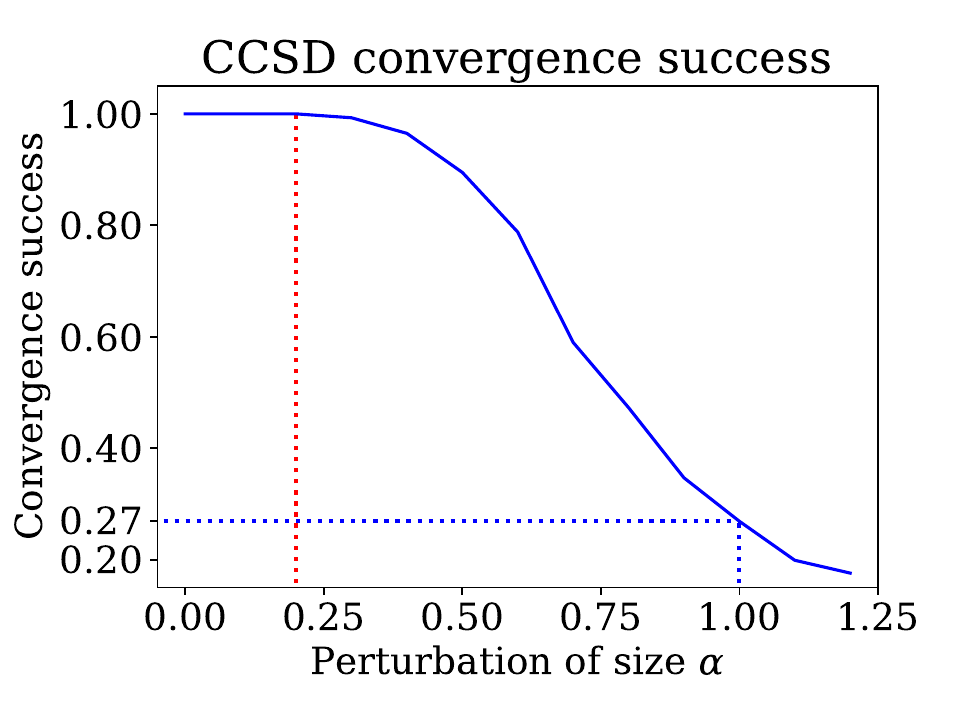}
    \caption{}
     \end{subfigure}
     \hfill
     \begin{subfigure}[t]{0.45\textwidth}
    \includegraphics[width = \textwidth]{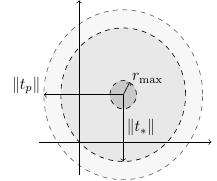}
     \caption{}
     \end{subfigure}
      \caption{\label{fig:CCBasinOfAtt}
      (a) The convergence success of 1000 simulations initializing CCSD with the optimal solution ${\bf t}_*$ plus a random perturbation ${\bf t}_p$ of size $\alpha = \Vert {\bf t}_p \Vert/\Vert {\bf t}_*\Vert$. The red dotted line indicates $r_{\rm max}$. The blue dotted line indicates the success rate within a ball around ${\bf t}_*$ of radius $\Vert {\bf t}_*\Vert$. (b)~Schematic representation of the different regions of convergences around ${\bf t}_*$.}
\end{figure}

In Figure~\ref{fig:CCBasinOfAtt}, we plot the success rate of Newton's method, i.e., how many of the randomly perturbed initializations converged toward ${\bf t}_*$, as a function of $\alpha
= \Vert {\bf t}_p \Vert/\Vert {\bf t}_*\Vert$. In particular, the initialization ${\bf t}_{\rm init} =0$ lies on the boundary of a ball around ${\bf t}_*$ of radius $\Vert {\bf t}_* \Vert$. Note that the (quasi) Newton iteration is commonly initialized by second-order M{\o}ller–Plesset amplitudes, however, using ${\bf t}_{\rm init} =0$ as initialization yields precisely those amplitudes in the first iteration. 

Figure~\ref{fig:CCBasinOfAtt} also shows that $r_{\rm max}$ is approximately $0.2 ~\Vert {\bf t}_* \Vert $. Moreover, it shows that beyond this point convergence towards ${\bf t}_*$ is by no means guaranteed. In fact, for an initialization that is placed randomly in a ball around  ${\bf t}_*$ of radius $\Vert {\bf t}_* \Vert$, the success rate is merely $27\%$. 
Being oblivious about the physical motivation of this initial guess, one could argue that it is quite surprising that Newton's method converges for the initial guess zero. 

In comparison, we find that the augmented Lagrangian always converges to the solution ${\bf t}_*$. 

\begin{figure}[h!]
    \centering
    \includegraphics[width = 0.5\textwidth]{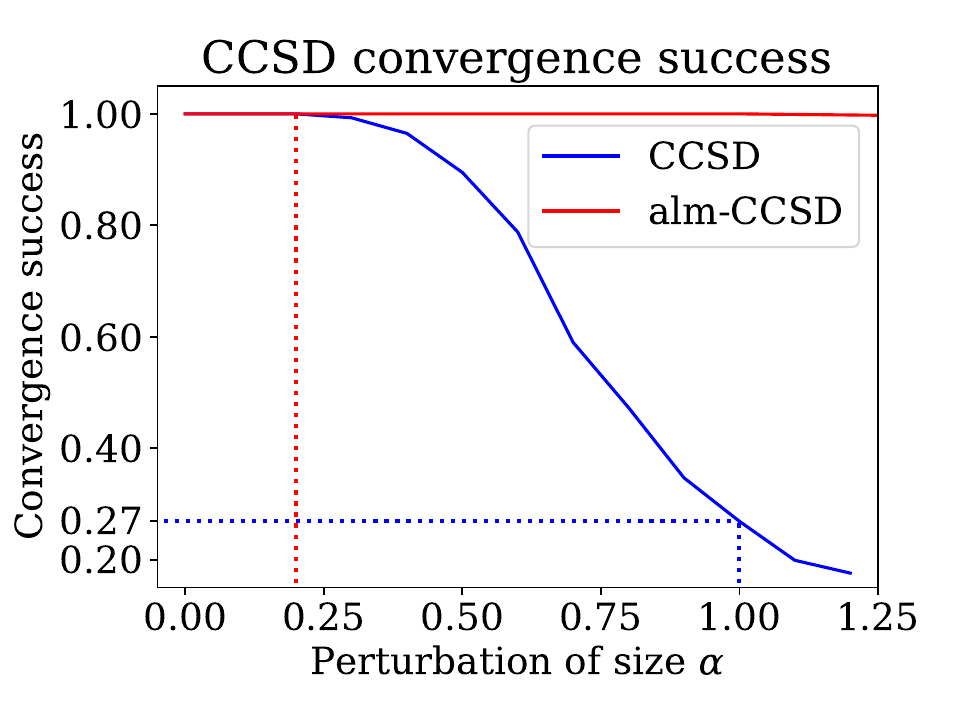}
    \caption{The convergence success of alm-CCSD and CCSD initialized with the optimal solution ${\bf t}_*$ plus a random perturbation ${\bf t}_p$ of size $\alpha = \Vert {\bf t}_p \Vert/\Vert {\bf t}_*\Vert$}
    \label{fig:aug-ccsd}
\end{figure}

\section{Conclusion}

In this article, we have presented a new optimization perspective on the coupled
cluster (CC) method. The CC method changes a linear high-dimensional
eigenvalue problem with exponential size into a low-dimensional system of
nonlinear system of equations. These equations are feasible to solve, but
converging to the root that represents the ground state has been challenging.
Our perspective suggests that an augmented Lagrangian formalism can help address
these issues without a significant performance overhead. 

We demonstrated the efficacy of our approach using the H$_4$ model system, a
staple in quantum chemistry for probing the nuances of CC methods. Our analysis
highlighted specific scenarios where the (quasi) Newton method tends to fail. In
these instances, we found that the CC method, optimized using an augmented
Lagrangian formulation, exhibits robust performance, effectively overcoming the
limitations observed with the conventional approach.

We therefore conclude that the augmented Lagrangian formulation presented in
this article represents a promising approach to the CC equations that is
reliable and efficient. This is particularly evident in scenarios where the root-finding (quasi) Newton method is prone to failure. As a result,
our approach provides a more versatile and general framework for applying the CC
method, broadening its realm of applicability in quantum chemistry.

\bibliographystyle{IEEEtranS}
\bibliography{lib}

\newpage
\section{Appendix}

\subsection{The Schrödinger equation}\label{section: background Seq}
The starting point of this article is the continuous {\em electronic Schrödinger equation}
\begin{equation} \label{eq:schroedinger}
\mathscr{H} \,\ket{w(\mathbf{x}_1,\mathbf{x}_2,\ldots ,\mathbf{x}_N)} \,\,=\,\, 
\lambda \,\ket{w(\mathbf{x}_1,\mathbf{x}_2 \ldots,\mathbf{x}_N)}.
\end{equation}
From this differential equation, the finite-dimensional eigenvalue problem mentioned in Eq.~\eqref{eq:eigenvector} can be derived, see e.g.~\cite{vcivzek1966correlation,szabo2012modern,helgaker2013molecular,faulstich2024recent,lehtola2020overview,faulstich2023algebraic}.
The unknown in Eq.~\eqref{eq:schroedinger} is
the eigenfunction $\ket{w(\mathbf{x}_1,\mathbf{x}_2,\ldots,\mathbf{x}_N)}$.
The arguments in this function are pairs $\mathbf{x}_i=(\mathbf{r}_i,s_i)$ with $\vr_i= (r_i^{(1)},r_i^{(2)},r_i^{(3)})$ in $\mathbb{R}^3$ and $s_i \in \{\pm 1/2 \}$ that represent the positions and spins of $N$ electrons. 
Since the $w$ describes the system's electrons which must adhere to {\em Pauli's exclusion principle}, the eigenfunction $\ket{w}$ must be antisymmetric in its $N$ arguments, i.e.~if ${\bf x}_i$ and ${\bf x}_j$ are switched then $\ket{w}$ is replaced by $-\ket{w}$.
The Hamiltonian $\mathscr{H}$ in Eq.~\eqref{eq:schroedinger} is a second-order differential operator which describes the behavior of $N$ interacting electrons in the vicinity of $N_{\rm nuc}$ stationary nuclei, i.e.,
\begin{equation}
\label{eq:ElecSE}
\mathscr{H}
\,\,=\,\,
-\,\frac{1}{2}\sum_{i=1}^N\Delta_{\vr_i}
\,-\,\sum_{i=1}^N\sum_{j=1}^{N_{\rm nuc}}\frac{Z_j}{\vert \vr_i-\vR_j\vert} 
\,+\,\sum_{i=1}^N\sum_{j=i+1}^N\frac{1}{\vert \vr_i-\vr_j\vert}.
\end{equation}
The symbol $\Delta_{\vr_i}$ in the leftmost sum denotes the Laplacian
$\, \sum_{j=1}^3 (\partial /\partial r_i^{(j)} )^2$.
All other summands in Eq.~\eqref{eq:ElecSE} act on $w$ by multiplication.
They contain constants which we now explain.
The atoms and their nuclei are indexed by $j=1,2,\ldots,N_{\rm nuc}$.
The constant $Z_j$ is the $j$th nuclear charge.
This is the atomic number listed in the periodic table, i.e. $Z_j$ is a positive integer.
The position of the $j$th nucleus is the point $\vR_j \in \mathbb{R}^3$ which is also constant.
We here consider only charge-neutral molecules.
This means that $N=\sum_{j=1}^{N_{\rm nuc}} Z_j$ and $N_{\rm nuc} \leq N$.

To illustrate the unfavorable scaling of the eigenvalue problem, we disregard the spin-degree of freedom for now, i.e., the eigenfunction $w$ is a function on $\mathbb{R}^{3N}$. Considering CO$_2$, which has 22 electrons, following a na\"ive grid discretization with 10 grid points per axis would then yield $10^{66}$ grid points. This kind of discretization rapidly becomes numerically intractable, yielding the development of multiple discretization techniques employed by the community, see~\cite{lehtola2020overview} for an overview. 

\subsection{Implementation details and spin restriction}

The here presented calculations were performed in the spin-restricted formulation. We shall now elaborate in detail on the respective spin symmetries in this formulation. 
In practice, $\chi_I({\bf r}, s)$ is product of two functions, i.e.,  $\xi_i:\mathbb{R}^3\to \mathbb{C}$ and $m_\gamma : \{\pm 1/2\} \to \{0,1\}$, where
\begin{equation}
m_0(s) \,=\, 
\left\lbrace
\begin{split}
1 \quad &{\rm if}~ s=+1/2,\\
0 \quad &{\rm if}~ s=-1/2,
\end{split}
\right.
\quad 
{\rm and}
\quad
m_1(s)\, =\, 
\left\lbrace
\begin{split}
1 \quad &{\rm if}~ s=-1/2,\\
0 \quad &{\rm if}~ s=+1/2.
\end{split}
\right.
\end{equation}
The binary functions $m_\gamma$ are spinors describing the spin -- a crucial degree of freedom in electronic structure theory that distinguishes physical states~\cite{yserentant2010regularity}. Without loss of generality, we assume the following ordering of $\mathcal{B}$:
\begin{equation}
\chi_{i+\gamma k}({\bf r}, s) \,\, = \,\,
\xi_i(\bf r)\,   m_\gamma(s), 
\end{equation}
i.e., we first list the $k$ function that correspond to $\gamma = 0$ (``spin 1/2'' functions) and then the $k$ function that correspond to $\gamma = 1$  (``spin -1/2'' functions). In order to distinguish more explicitly between  $\gamma = 0$ and  $\gamma = 1$ functions, we may employ the notation $\chi_{i,\gamma }$. 

The computations presented here use the \texttt{PySCF} software package~\cite{sun2020recent,sun2018pyscf,sun2015libcint} for the computations of $h$ and $v$ in Eq.~\eqref{eq:ElecHam}, as well as for the out-of-the-box benchmark computations. 
The subsequently presented computations are in the {\it spin-restricted formulation} which we shall elaborate on in more detail. 
As outlined above, the electronic spin is an intrinsic quantity that -- although not explicitly appearing in the Hamiltonian -- affects its eigenstates~\cite{yserentant2010regularity}.
Mathematically, the effect that different spin symmetries (such as the spin-restricted formulation) have on the computations is imposing certain symmetries onto quantities like $h$, $v$, $T$, and $\Lambda$, see Sec.~\ref{sec:Background}. 

In the spin-restricted formulation, the single particle part of the Hamiltonian $H$ takes the form
\begin{equation}
h
=
\sum_{\substack{(p,\gamma)\\(q, \gamma)}}
h_{p,q} a^\dagger_{a \gamma} a_{i \gamma},
\end{equation}
in particular, $h$ is block-diagonal with only two blocks on the diagonal that are identical. \\
The single excitation term in the cluster matrix reads
\begin{equation}
T_1
= 
\sum_{\substack{(i,\gamma)\\(a, \gamma)}}
t_{{a\gamma \choose i\gamma}}
a^\dagger_{a \gamma} a_{i \gamma}
\end{equation}
where the spin-restricted formulation enforces that 
\begin{equation}
t_{{a 0 \choose i 0}}
=
t_{{a 1 \choose i 1}}.
\end{equation}
Hence, similar to the single particle part of $H$, the single excitation part in $T$ is block-diagonal with only two blocks on the diagonal that are identical.\\
The double excitation part in $T$ is slightly more involved. It reads 
\begin{equation}
\label{eq:T2}
T_2 
= \frac{1}{4}\sum_{IJAB} t_{{AB \choose IJ}} a_{A}^{\dagger} a_{I} a_{B}^{\dagger} a_{J}
= \sum_{\substack{I<J,\\ A<B}} t_{{AB \choose IJ}} a_{A}^{\dagger} a_{I} a_{B}^{\dagger} a_{J},
\end{equation}
where $I,J,A,B$ include both the spatial and spin indices. 
The two forms are equivalent because $t_{{AB \choose IJ}}$ is antisymmetric w.r.t. both $IJ$ and $AB$. 
In order to compute the derivative in Eq.~\eqref{eq:derivativeT} correctly, we have to take into account that $t_{{AB \choose IJ}}$ are not independent variables because of 1) the antisymmetry constraints and 2) the spin restrictions. 
To eliminate dependent variables, we can reparametrize $T$. 
First, we replace $I,J, A,B$ with the explicitly spin-dependent indices, i.e.,
\begin{equation}
T_2 
= \sum_{\substack{(i,\sigma)<(j, \sigma')\\ (a,\gamma)<(b, \gamma')}}  
t_{{a \gamma b \gamma' \choose i\sigma j\sigma'}} 
a_{a \gamma}^{\dagger} 
a_{i \sigma} 
a_{b \gamma'}^{\dagger} 
a_{j \sigma'} 
\delta_{\sigma+\sigma', \gamma+\gamma'}.
\end{equation}
As outlined above, we encode spin $1/2$ as $\mu = 0$ and spin $-1/2$ as $\mu = 1$ so that $\sigma + \sigma'$ is meaningful, and the delta function ensures that the excitation conserves the spin. 
The meaning of ``$<$'' can in principle be any order, we here choose it to be $(i, \sigma) < (j, \sigma')$ if and only if $\sigma < \sigma' \lor (\sigma = \sigma' \land i < j)$. 
The sum $\sum_{I<J}$ can be decomposed into a ``same spin part'' and an ``opposite spin part'', i.e., 
\begin{equation}
    \sum_{(i, \sigma) < (j, \sigma')} = \sum_{i<j} \sum_{\sigma, \sigma'} \delta_{\sigma,\sigma'} + \sum_{ij} \sum_{\sigma, \sigma'} \delta_{\sigma, 0} \delta_{\sigma, 1}.
\end{equation}
We can do the same for the other sum $\sum_{A < B}$ and multiply the expressions. The cross terms of this multiplication vanish because they do not conserve the spin. 
The result consists of a ``same spin excitation'' and an ``opposite spin excitation'', i.e., 
\begin{equation}
\sum_{\substack{I < J\\A < B}}
= 
\sum_{\substack{\sigma,\sigma'\\\gamma, \gamma'}}
\left(
\sum_{\substack{i<j\\a<b}} 
\delta_{\sigma, \sigma'} \delta_{\gamma, \gamma'}
+ \sum_{ijab} 
\delta_{\sigma, 0}  \delta_{\sigma', 1}  \delta_{\gamma, 0} \delta_{\gamma', 1}
\right).
\end{equation}
Applying this to Eq.~\eqref{eq:T2} and contracting the delta terms yields
\begin{equation}
    T =  \sum_{\substack{i<j\\a<b}} \sum_{\sigma} 
    t_{{a \sigma b \sigma \choose i \sigma j\sigma}}
    a_{a \sigma}^{\dagger} a_{i \sigma} a_{b \sigma}^{\dagger} a_{j \sigma}
    + \sum_{ijab} t_{{a1 b0 \choose i1 j0}}
    a_{a 1}^{\dagger} a_{i 1} a_{b 0}^{\dagger} a_{j 0}.
\end{equation}
The spin-restricted formulation imposes further restrictions, namely,
\begin{equation}
\begin{aligned}
\sum_{\substack{i\neq j\\ a \neq b}}
\sum_{\sigma} 
\frac{1}{2} 
t_{{ab \choose ij}} a_{a \sigma}^{\dagger}
a_{i \sigma} a_{b \sigma}^{\dagger} a_{j \sigma} &= 
\sum_{\substack{i < j\\ a \neq b}}
\sum_{\sigma} 
\frac{1}{2} 
\left(
t_{{ab \choose ij}} 
a_{a \sigma}^{\dagger}
a_{i \sigma} 
a_{b \sigma}^{\dagger} 
a_{j \sigma}
+
t_{{ab \choose ij}} 
a_{a \sigma}^{\dagger}
a_{j \sigma} 
a_{b \sigma}^{\dagger} 
a_{i \sigma}\right) \\
&=  
\sum_{\substack{i < j\\ a \neq b}}
\sum_{\sigma} 
\frac{1}{2} (t_{{ab \choose ij}} - t_{{ab \choose ji}})
a_{a \sigma}^{\dagger}
a_{i \sigma} 
a_{b \sigma}^{\dagger} 
a_{j \sigma}\\
&=  
\sum_{\substack{i < j\\ a < b}}
\sum_{\sigma} 
\frac{1}{2} \left(
(t_{{ab \choose ij}} - t_{{ab \choose ji}})
-(t_{{ba \choose ij}} - t_{{ba \choose ji}})
\right)
a_{a \sigma}^{\dagger}
a_{i \sigma} 
a_{b \sigma}^{\dagger} 
a_{j \sigma}.\\
\end{aligned}
\end{equation}
Therefore, the constraint has to be
\begin{equation}
t_{i\sigma j\sigma \choose a \sigma b\sigma} 
= 
\frac{1}{2} 
\left(
t_{{ab \choose ij}} + t_{{ba \choose ji}} - t_{{ab \choose ji}} - t_{{ba \choose ij}}
\right)
\end{equation}
The above derivation led to a formulation where $t_{{ab \choose ij}}$ are now independent variables, for which we can seek the stationary condition.
This yields
\begin{equation}
\frac{\partial T}{\partial t_{ij \choose ab}}
=  
a_{a 0}^{\dagger} 
a_{i 0} 
a_{b 1}^{\dagger} 
a_{j 1}
+ 
\frac{1}{2} \sum_{\sigma} (1 -\delta_{ij})(1 - \delta_{ab})
a_{a \sigma}^{\dagger} 
a_{i \sigma} 
a_{b \sigma}^{\dagger} 
a_{j \sigma},
\end{equation}
and finally
\begin{equation}
T_2
= 
\sum_{ijab} \frac{\partial T}{\partial t_{ij \choose ab}} t_{ij \choose ab}.
\end{equation}

\subsection{Proof of Proposition~\ref{th:Commutator}}
\begin{proof}
Let 
$$
X_\mu = X_{a_1,...,a_k \choose i_1,...,i_k}
=
a_{a_k}^\dagger ... a_{a_1}^\dagger a_{i_k} ... a_{i_1}
\quad \text{and} \quad
X_\nu = X_{b_1,...,b_\ell \choose j_1,...,j_\ell}
=
a_{b_\ell}^\dagger ... a_{b_1}^\dagger a_{j_\ell} ... a_{j_1}.
$$
The proof is conducted in two steps:\\

First, we seek to permute all creation operators to the left using the CAR.
We begin with the following product and note that when permuting $a_{b_\ell}^\dagger$ to the right of $a_{a_1}^\dagger$ we merely pick up a sign, since $b_\ell \notin v_{\rm occ}$, i.e., 
$$
\wick{
a_{a_k}^\dagger ... a_{a_1}^\dagger \c1 a_{i_k} ... a_{i_1}
\c1 a_{b_\ell}^\dagger ... a_{b_1}^\dagger a_{j_\ell} ... a_{j_1}
}
=
(-1)^{k}
a_{a_k}^\dagger ... a_{a_1}^\dagger a_{b_{\ell}}^\dagger a_{i_k} ... a_{i_1}
a_{b_{\ell-1}}^\dagger ... a_{b_1}^\dagger a_{j_\ell} ... a_{j_1}
$$
This furthermore yields
$$
a_{a_k}^\dagger ... a_{a_1}^\dagger  a_{i_k} ... a_{i_1}
 a_{b_\ell}^\dagger ... a_{b_1}^\dagger a_{j_\ell} ... a_{j_1}
=
(-1)^{\ell \cdot k}
a_{a_k}^\dagger ... a_{a_1}^\dagger a_{b_{\ell}}^\dagger ... a_{b_1}^\dagger a_{i_k} ... a_{i_1}
a_{j_\ell} ... a_{j_1}
$$
and similar
$$
a_{b_\ell}^\dagger ... a_{b_1}^\dagger a_{j_\ell} ... a_{j_1}
 a_{a_k}^\dagger ... a_{a_1}^\dagger  a_{i_k} ... a_{i_1}
=
(-1)^{\ell \cdot k}
a_{b_\ell}^\dagger ... a_{b_1}^\dagger a_{a_k}^\dagger ... a_{a_1}^\dagger
a_{j_\ell} ... a_{j_1} a_{i_k} ... a_{i_1}
$$

Second, we wish to unify the index sequence of the creation and annihilation operators respectively. 
Applying the CAR again, we find
$$
\wick{
\c1 a_{b_\ell}^\dagger ... a_{b_1}^\dagger  \c1 a_{a_k}^\dagger ... a_{a_1}^\dagger
a_{j_k} ... a_{j_1} a_{i_k} ... a_{i_1}
}
=
(-1)^{\ell }
a_{a_k}^\dagger a_{b_\ell}^\dagger ... a_{b_1}^\dagger   a_{a_{k-1}}^\dagger ... a_{a_1}^\dagger
a_{j_k} ... a_{j_1} a_{i_k} ... a_{i_1}
$$
which yields
$$
a_{b_\ell}^\dagger ... a_{b_1}^\dagger  a_{a_k}^\dagger ... a_{a_1}^\dagger
a_{j_\ell} ... a_{j_1} a_{i_k} ... a_{i_1}
=
(-1)^{2*\ell \cdot k}
a_{a_k}^\dagger ... a_{a_1}^\dagger
a_{b_\ell}^\dagger ... a_{b_1}^\dagger    
a_{i_k} ... a_{i_1}
a_{j_\ell} ... a_{j_1}
$$
overall this yields
\begin{equation}
\begin{aligned}
[X_\mu, X_\nu] 
&= 
[X_{a_1,...,a_k \choose i_1,...,i_k}, X_{b_1,...,b_\ell \choose j_1,...,j_\ell} ]\\
&=
a_{a_k}^\dagger ... a_{a_1}^\dagger a_{i_k} ... a_{i_1}
 a_{b_\ell}^\dagger ... a_{b_1}^\dagger a_{j_\ell} ... a_{j_1}\\
&\quad -
a_{b_\ell}^\dagger ... a_{b_1}^\dagger a_{j_\ell} ... a_{j_1}
 a_{a_k}^\dagger ... a_{a_1}^\dagger  a_{i_k} ... a_{i_1}\\
&=
(-1)^{\ell \cdot k}
a_{a_k}^\dagger ... a_{a_1}^\dagger a_{b_{\ell}}^\dagger ... a_{b_1}^\dagger a_{i_k} ... a_{i_1}
a_{j_\ell} ... a_{j_1}\\
&\quad- (-1)^{\ell \cdot k}
a_{b_\ell}^\dagger ... a_{b_1}^\dagger a_{a_k}^\dagger ... a_{a_1}^\dagger
a_{j_\ell} ... a_{j_1} a_{i_k} ... a_{i_1}\\
&=
(-1)^{\ell \cdot k}
a_{a_k}^\dagger ... a_{a_1}^\dagger a_{b_{\ell}}^\dagger ... a_{b_1}^\dagger a_{i_k} ... a_{i_1}
a_{j_\ell} ... a_{j_1}\\
&\quad- 
(-1)^{3*\ell \cdot k}
a_{a_k}^\dagger ... a_{a_1}^\dagger
a_{b_\ell}^\dagger ... a_{b_1}^\dagger    
a_{i_k} ... a_{i_1}
a_{j_\ell} ... a_{j_1}\\
&=0.
\end{aligned}
\end{equation}
\end{proof}

\subsection{Proof of Proposition~\ref{th:nilpotency}}

\begin{proof}
This follows from $(a_p^\dagger)^2 = (a_p)^2 = 0$.
Let 
$$
X_\mu = X_{a_1,...,a_k \choose i_1,...,i_k}
=
a_{a_k}^\dagger ... a_{a_1}^\dagger a_{i_k} ... a_{i_1}.
$$
Then
\begin{equation}
\begin{aligned}
X_\mu^2
= 
\wick{
\c1 a_{a_k}^\dagger ... a_{a_1}^\dagger a_{i_k} ... a_{i_1} \c1 a_{a_k}^\dagger ... a_{a_1}^\dagger a_{i_k} ... a_{i_1}
}
=
- a_{a_k}^\dagger a_{a_k}^\dagger ... a_{a_1}^\dagger a_{i_k} ... a_{i_1}  a_{a_k-1}^\dagger ... a_{a_1}^\dagger a_{i_k} ... a_{i_1}
=
0
\end{aligned}
\end{equation}
\end{proof}

\subsection{Gradients}

The gradient of the energy and the constraints can be derived by looking at the first order change
\begin{equation}
\begin{aligned}
    E(t + \delta t) - E(t) &=  \bra{\phi_0} H e^{T(t) + T(\delta t)} \ket{\phi_0}
    - \bra{\phi_0} H e^{T(t)} \ket{\phi_0}\\
    &=   \bra{\phi_0}  H e^{T(t)} (I + T(\delta t)) \ket{\phi_0}
    - \bra{\phi_0} H e^{T(t)} \ket{\phi_0}\\
    &= \bra{\phi_0} H e^{T(t)} T(\delta t) \ket{\phi_0}.
\end{aligned}
\end{equation}
\begin{equation}
\begin{aligned}
    g_{\mu}(t + \delta t) - g_{\mu}(t) &= \bra{ \phi_{\mu}} e^{-T(t)}(1 -T(\delta t)) H e^{T(t)}
    (1+T(\delta t)) \ket{\phi_0} - \bra{\phi_{\mu}} e^{-T(t)} H e^{T(t)} \ket{\phi_0}\\
    &= \bra{ \phi_{\mu}} e^{-T(t)} H T(\delta t) e^{T(t)} \ket{\phi_0}
    - \bra{ \phi_{\mu}} e^{-T(t)} T(\delta t) H e^{T(t)} \ket{\phi_0}\\
    &= \bra{\phi_{\mu}}  e^{-T(t)} [H, T(\delta t)] e^{T(t)} \ket{\phi_0}.
\end{aligned}
\end{equation}
The result is
\begin{align}
\frac{\delta E}{\delta t_{\mu}}  = 
\bra{\phi_0} H e^{T(t)} X_{\mu} \ket{\phi_0},\quad
\frac{\delta g_{\mu}}{\delta t_{\nu}}  =  
\bra{ \phi_{\mu}} e^{-T(t)} [H, X_{\nu}] e^{T(t)} \ket{\phi_0},\quad
X_{\mu} = \frac{T(\delta t)}{\delta t_{\mu}}.
\end{align}
\end{document}